# Wide-open, high-resolution microwave/millimeter-wave Doppler frequency shift estimation using photonics technology

Xihua Zou [1,*], Wangzhe Li [2,*], Bing Lu [1], Wei Pan [1], Lianshan Yan [1], and Liyang Shao [1]

**Abstract** Today, wide-open, high-resolution Doppler frequency shift (DFS) estimation is essential for radar, microwave/millimeter-wave, and communication systems. Using photonics technology, an effective approach is proposed and experimentally demonstrated, providing a high-resolution and frequency-independent solution. In the approach consisting of two cascaded opto-electronic modulators, DFS between the transmitted microwave/ millimeter-wave signal and the received echo signal is mapped into a doubled spacing between two target optical sidebands. Subsequently, the DFS is then estimated through the spectrum analysis of a generated low-frequency electrical signal, with an improved resolution by a factor of 2. In experiments, DFSs from -90 to 90 KHz are successfully estimated for microwave/millimeter-wave signals at 10, 15, and 30 GHz, where estimation errors keep lower than $\pm 5\times 10^{-10}$ Hz. For radial velocity measurement, these results reveal a range from 0 to 900 m/s (0 to 450 m/s) and a resolution of $1\times 10^{-11}$ m/s ($5\times 10^{-12}$ m/s) at 15-GHz (30-GHz) frequency band.

## 1. Introduction

Doppler effect arising from the relative motion between the observer and the wave source, has been extensively demonstrated or/and used in wireless communications, scientific measurement, electronic warfare, and radar applications [1-3]. In these scenarios, the key task is to quantitatively and accurately discriminate the corresponding Doppler frequency shift (DFS). Conventional electrical approaches to DFS estimation include for instance the direct Fourier analysis, the joint time-frequency analysis, the inphase-quadrature (I-Q) mixing setup [1-3], and so on. Recently, there are already several big challenges facing the estimation and analysis of DFS. Firstly, fine resolution is urgently required, to improve the performances such as the capability of detecting low-speed moving objects and the accuracy of precious velocity detection. More importantly, since radar and communication systems operating at a large variety of frequency bands from MHz to hundreds of GHz [2, 4, 5] are in service, it is essentially expected that the estimation should be independent of the microwave/millimeter-wave frequency or be effectively implemented within a large wide-open frequency range, which may be difficult to be fulfilled by using a single electrical approach or device. In addition, as for the high-velocity scenarios (e.g., satellite, missile) and millimeter-wave applications, a large measurement range from several Hz to tens/hundreds of KHz [2], or even MHz is typically required for DFS estimation.

Thanks to the intrinsic features of photonics technology in wide instantaneous bandwidth and immunity to electro-magnetic interference, photonic or photonic-assisted approaches and setups are emerging as promising and powerful solutions for the generation, manipulation, transmission, receiving, processing, analysis and measurement of wideband, complex microwave/millimeter-wave signals [6-18], particularly suitable for defence applications [19-21]. Here, we focus on the advances and the realization of photonic microwave/millimeter-wave signal analysis and measurement. Some examples are given in the following. For spectrum analysis, a bandwidth up to THz [22] or a resolution of 3.4 MHz [23], has demonstrated in photonic-assisted approaches. For instantaneous frequency measurement, related photonic approaches were designed by using dispersion-induced microwave power fading [24-27], time delay-based phase discrimination [28-34], and photonic-assisted channelization [35-39]. The arrival of angle (AOA) and the time-difference-of-arrival (TDOA) can also be discriminated in the optical domain [40-43]. In [41], spatial-spectral (S2) materials were employed to perform frequency hole burning, and the AOA was estimated from the analysis of the absorption spectrum. In [42] and [43], two electro-optic modulators connected in parallel or in series were employed, wherein the AOA information was equivalently converted into the optical power variations. Also, an approach based on a dual-parallel Mach-Zehnder modulator was reported for AOA measurement [44]. According to these approaches and methods, it is clear that photonics enables us to enrich the functionalities and to enhance the performances for the analysis and measurement of microwave/ millimeter-wave signals. Thus, as one of significant functionalities for microwave/ millimeter-wave signal measurement and analysis, it is really anticipated that DFS estimation can benefit from photonics technology, to improve the resolution and to enlarge the wide-open frequency range. However, to the best of our knowledge, no effective method has been reported to date to realize this functionality (i.e., DFS estimation) with specific performance requirements mentioned above.

In this paper, an effective approach using photonics technology is proposed and demonstrated experimentally to measure DFSs, which might be the first case regarding photonic or photonic-assisted DFS estimation capable of providing finer resolution and wide-open frequency range. In the proposed approach, two electro-optic modulators (i.e., EOM-I and EOM-II) that are connected in series, the first one biased at the minimum transmission point to suppress optical carrier and the second at the quadrature point. A replica of the transmitted microwave/ millimeter-wave signal is applied to EOM-I, while the received echo signal with an unknown DFS is launched into EOM-II. At the output of EOM-II, two target optical sidebands close to the optical carrier from the laser diode are generated, of which the frequency spacing is just twice the DFS to be estimated. A frequency beating between the two target optical sidebands in a low-speed photodetector (PD) yields a low-frequency electrical signal, and the DFS of interest can be obtained through spectrum analysis of this signal. It is noted that a twofold improvement in the estimation resolution is obtained, since a doubled frequency spacing has been formed and equivalently measured in the optical domain. Moreover, the DFS estimation is totally independent of the microwave/millimeter-wave frequency, eliminating the limit on the frequency coverage in a single electrical approach. Said other words, the proposed approach is guaranteed wide-open operation.

---
[1] Center for Information Photonics and Communications, School of Information Science and Technology, Southwest Jiaotong University, Chengdu, 610031, China.
[2] Microwave Photonics Research Laboratory, University of Ottawa, Ottawa, ON, K1N 6N5, Canada.
* Corresponding authors: zouxihua@swjtu.edu.cn, wli008@uottawa.ca.

## 2. Operation principle

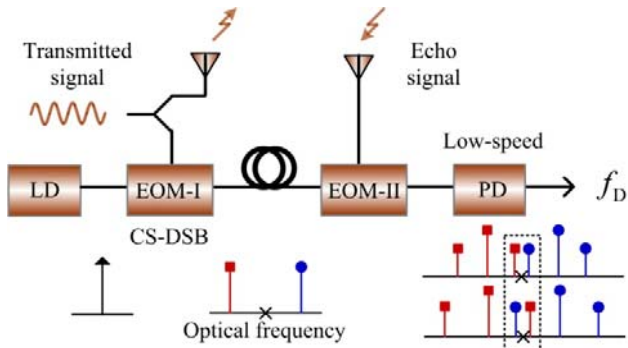

Fig. 1 Schematic diagram of the proposed approach for Doppler frequency shift estimation. (LD, laser diode; EOM, electro-optic modulator; CS-DSB, carrier-suppressed double sideband modulation; PD, photodetector)

The estimation approach is schematically illustrated in Fig. 1, which consists of a CW laser diode (LD), EOM-I and EOM-II cascaded in series, and a low-speed PD. A replica of the transmitted microwave/millimeter-wave signal is applied to EOM-I biased at the minimum transmission point to externally modulate the optical carrier from the LD, providing carrier-suppressed double sideband (CS-DSB) modulation format. Thus, two optical sidebands (i.e., the upper and the lower sidebands) are generated, as

$$E_1(t) = J_{-1}(\beta_1)\exp j[2\pi(f_0 - f_m)t] + J_{+1}(\beta_1)\exp j[2\pi(f_0 + f_m)t] \quad , \quad (1)$$

where $f_0$ (or $\omega_0 = 2\pi f_0$) is the frequency (or the angular frequency) of the optical carrier, $f_m$ is the frequency of the transmitted microwave/millimeter-wave signal, $\beta_1$ is the modulation index in EOM-I, and $J_{\pm 1}()$ represents the ±1st Bessel function of the first kind.

The two sidebands in Eq. (1) are then injected into EOM-2 where they serve as two optical carriers. When an echo signal reflected from a moving object is received, a relative frequency shift occurs between the transmitted and the echo signals due to Doppler effect. Applying the echo signal to EOM-II yields two target optical sidebands located close to the optical carrier of the LD, of which one is inducted by the upper sideband and the other by the lower sideband. As such, the two target optical sidebands can be written as

$$E_2(t) = J_{-1}(\beta_1)J_{+1}(\beta_2)\exp j[2\pi(f_0 - f_m + f'_m)t] + J_{+1}(\beta_1)J_{-1}(\beta_2)\exp j[2\pi(f_0 + f_m - f'_m)t] \quad , \quad (2)$$

where $\beta_2$ is the modulation index in EOM-2, and $f'_m$ is the frequency of the received echo signal. From the inset in Fig. 1 and Eq. (2), it is clearly seen that the DFS between the transmitted signal and the received echo signal is converted into the frequency spacing between the two target optical sidebands in the optical domain. A low-speed PD is used to convert the two target optical sidebands into a low-frequency electrical signal, since the DFSs in most cases are usually lower than 1 MHz (see Fig. 3.2 in [2]). Other high-frequency microwave signals originating from frequency beating are eliminated due to the bandwidth limit of the low-speed PD. Thus, the recovered electrical signal and the measured frequency spacing can be expressed as

$$E_m \propto \exp j(\Delta ft) \quad , \quad (3)$$

$$\Delta f = 2|f_m - f'_m| = 2f_D \quad , \quad (4)$$

where $\Delta f$ and $f_D$ are the frequency spacing between the two target optical sidebands and the DFS, respectively. Here the frequency shift may be $f_m - f'_m$ or $f'_m - f_m$, as illustrated in the inset of Fig. 1. Consequently, we are capable of estimating the DFS from the spectrum analysis of the low-frequency electrical signal in Eq. (3).

Next, from the measured DFS, the corresponding radial velocity of a moving object can be derived accordingly. For instance, in a monostatic radar system where the transmitter and the receiver are located in the same place, the radial velocity along the line of sight (LOS) direction can be calculated as

$$v = \frac{f_D}{2f_m}c \quad , \quad (5)$$

where $c$ is the light velocity in vacuum, and $v$ is the radial velocity of the moving object [2].

By substituting Eq. (4) into Eq. (5), the radial velocity can be directly calculated by following Eq. (6).

$$v = \frac{\Delta f}{4f_m}c \quad (6)$$

According to the principle above, the proposed approach exhibits several remarkable advantages. First of all, the estimation is implemented in the optical domain and is totally independent of the carrier frequency of the transmitted signal (i.e., signal source). Here an equivalent optical mixing or down-conversion is performed, and no high-frequency or ultra-wideband electrical mixing is required. Thus, a huge wide-open frequency range is resulted, which makes this proposed approach suitable for DFS estimation in L-, S-, C-, X-, Ku-, K-, Ka-, V-, and W-bands [5]. Additionally, due to the estimation procedure in the optical domain, the proposed approach is more robust to strong electro-magnetic interference. More importantly, from Fig. 1 and Eq. (4), the measured frequency spacing is twice the DFS under test, rather than the same value in conventional approaches. Subsequently, a doubled frequency value is obtained, and this information enables us to have the resolution of the DFS estimation and the radical velocity measurement enhanced by a factor of 2.

## 3. Realization and experiments

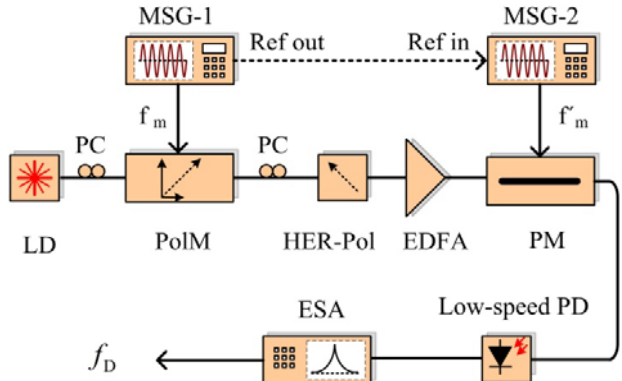

Fig. 2 Experimental setup of the proposed approach. (LD, laser diode; PC, polarization controller; PolM, polarization modulator; MSG, microwave signal generator; HER-Pol: high-extinction-ratio polarizer; EDFA, Er-doped fiber amplifier; PM, phase modulator; PD, photodetector; ESA, electrical spectrum analyzer)

To validate the proposed approach, proof-of-concept experiments are performed in this section. As outlined in Fig. 2, a polarization modulator (PolM, 40Gb/s) in combination with a high-extinction-ratio polarizer (HER-Pol) is employed to perform CS-DSB modulation, of which the details can be found in [45]. Here, an extinction ratio greater than 28 dB is provided to suppress the optical carrier as low as possible. Next, a phase modulator (PM) serves as EOM-II, providing merits of low half-wave voltage and no drift of bias control over an intensity modulator (IM). To emulate the DFS between the transmitted and the echo signals, two microwave signal generators (MSGs, Anritsu 3694B) are used. Since the reference output of one MSG is launched as the reference input of the other, the two microwave/millimeter-wave signals generated are triggered by the same reference signal and phase-correlated. By setting a frequency difference between the two microwave signals, an equivalent DFS is formed between the transmitted signal from MSG-1 and the echo signal from MSG-2, like the phenomenon in a CW Doppler radar.

To begin, two microwave signals at 15 and 15.001 GHz are generated, serving as the transmitted and the echo signals respectively. Under CS-DSB modulation, the optical carrier at 1548.575 nm from the LD is externally modulated by the transmitted signal at 15 GHz, two optical sidebands (i.e., the lower and the upper sidebands) being generated. As shown in Fig. 3a, a carrier suppression ratio larger than 30 dB is obtained, owing to the use of an HER-Pol. After a second modulation in the PM, two target optical sidebands around the original wavelength of 1548.575 nm are formed, which are induced by the lower sideband and the upper one in Fig. 3a, respectively, as illustrated in Fig. 3b and the inset of Fig. 1. Due to the limit on the resolution of the optical spectrum analyzer, the two target optical sidebands with a frequency spacing as small as 2 MHz between them cannot be differentiated from each other in Fig. 3b.

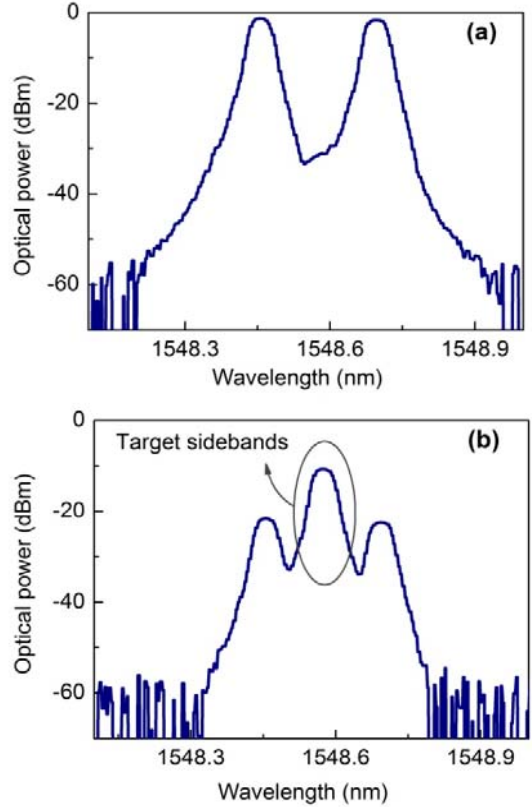

Fig. 3 Measured optical spectra of (a) the CS-DSB modulation and (b) the two target optical sidebands.

The two target optical sidebands are then sent to the low-speed PD for recovering low-frequency electrical signal, and the resulting spectrum is measured and presented in Fig. 4. It is obvious that a low-frequency component at 2 MHz occurs, indicating the 1-MHz DFS (i.e., the frequency difference between 15 and 15.001 GHz). Through the spectrum analysis of this frequency-doubled low-frequency electrical component, the DFS estimation is realized with a twofold improvement in the resolution. On the other hand, in addition to the frequency-doubled component, a fundamental component at 1 MHz is also observed in Fig. 4, which originates from the frequency beating between the residual optical carrier from a non-ideal CS-DSB modulation and the two target sidebands. Fortunately, a suppression ratio up to 40 dB is obtained between the frequency-doubled and the fundamental electrical components. Thus the fundamental component exerts very little influence on DFS estimation, which is omitted in the remaining of this paper.

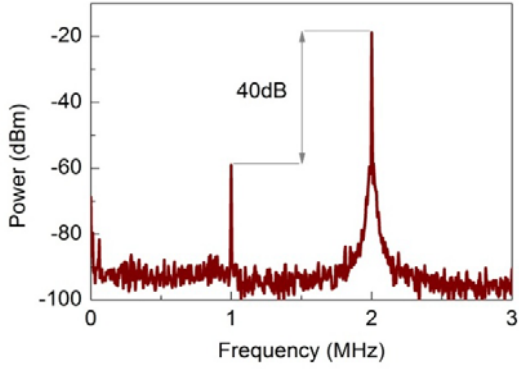

Fig. 4 Measured electrical spectrum at the output of the low-speed PD.

Likewise, different DFSs can be emulated by tuning the output frequency of MSG-2 away from 15 GHz in steps of 1, 10, and 100 KHz, while the output frequency of MSG-1 keeps fixed at 15 GHz. With the step of 1 KHz, the electrical spectrum for each frequency offset is recorded within a 2-KHz span by using an electrical spectra analyzer operating with a 10-Hz resolution bandwidth (RBW) and a 10-Hz video bandwidth (VBW), as illustrated in Fig. 5a. The DFSs are then extracted from these measured electrical spectra. As shown in Fig. 5b, the measured shifts are in line with the theoretical estimation in Eq. (1). Here, a doubled variation slope is formed, as compared with a conventional slope of 1 (see the bold solid line). In particular, for the frequency offset of zero, the estimation cannot be performed because of the existing of large DC distortions. So the DFS at the zero frequency offset and the corresponding estimation error are omitted, which might be reasonable for most practical applications where a non-zero frequency shift occurs.

According to the estimated DFSs, the estimation errors are further analyzed. As presented in Fig. 5c, resultant errors to the DFSs measurement keep lower than $\pm 5\times 10^{-10}$ Hz within the range from -9 to 9 KHz. It should be pointed out that the estimation errors or resolutions are associated with the performance of the used ESA to some extent. Lower errors or higher resolutions can be offered by using a high-performance ESA, and vice verse. Fortunately, there are already such commercially available ESAs with extremely fine resolution in low-frequency range. Furthermore, for a given resolution of the used ESA, the estimation resolution is improved by a factor of 2, compared with that of $\pm 1\times 10^{-9}$ Hz in conventional approaches. Because the DFS is mapped into a doubled frequency spacing in the optical domain. During the analysis on the estimation errors, the frequency accuracy of the MSGs (Anritsu 3694B) is the same as the internal or external 10-MHz time base under the CW mode, and their frequency instability might be canceled because a relative frequency offset between the two common-triggered and phase-correlated MSGs is used.

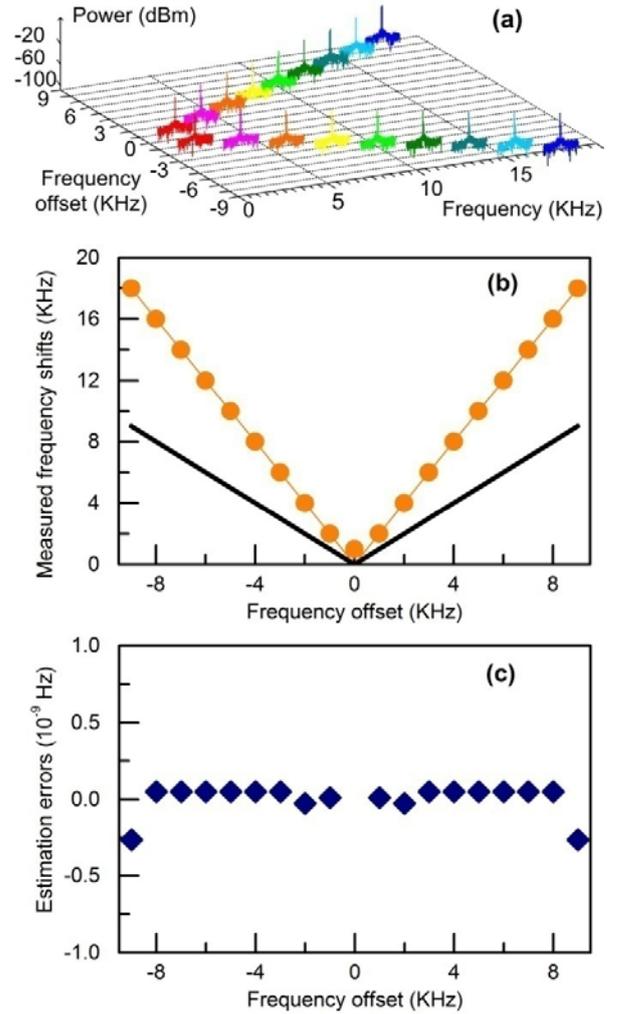

Fig. 5 Measured spectra (a), measured Doppler frequency shifts (b), and estimation errors (c) for the 15-GHz transmitted signal being tuned at an offset step of 1 KHz.

From these measured DFSs and the resulting estimation errors, the corresponding radial velocity of a moving object can be derived by using Eqs. (5) and (6). An estimation range from 0 to 90 m/s can be offered for the radial velocity measurement, due to a maximum DFS up to 9 KHz. Moreover, a resolution close to $1\times 10^{-11}$ m/s for radial velocity measurement is resulted from the obtained estimation errors or resolutions lower than $\pm 5\times 10^{-10}$ Hz in DFS estimation.

For other frequency offset steps of 10 and 100 KHz, the DFSs are then estimated and the resultant errors are analyzed as well. As shown in Fig. 6, with the same experimental setup for DFS estimation, the estimation errors are lower than $\pm 5\times 10^{-10}$ and $\pm 5\times 10^{-9}$ Hz, respectively. Note that these estimation errors are also half those of the conventional approaches, within a given measurement range. A comparison among these results in Fig. 6 shows that the estimation errors increase with the extension of the measurement ranges as well as the DFSs, indicating a trend in line with the traditional tradeoff between the measurement range and the estimation error or resolution.

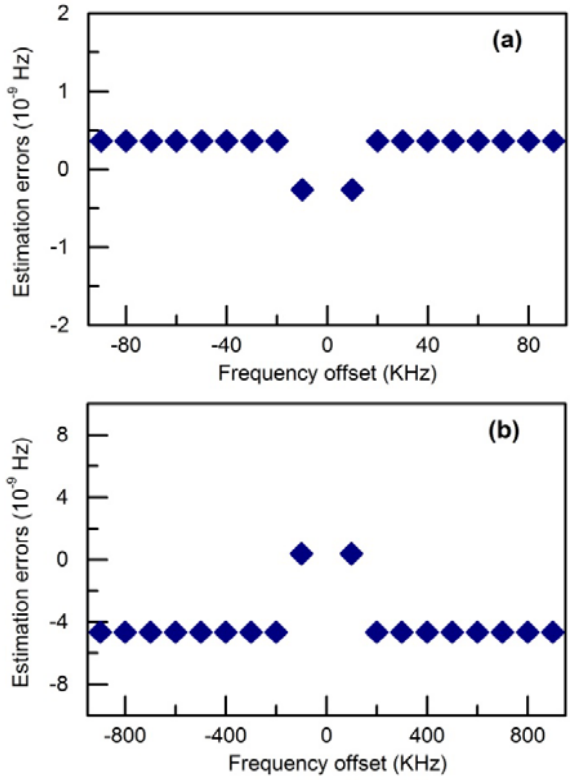

Fig. 6 Estimation errors for the 15-GHz transmitted signal being tuned at offset steps of (a) 10 KHz and (b) 100 KHz.

Similarly, a radial velocity from 0 to 900 m/s can be covered with a resolution of $1\times10^{-11}$ m/s for the 10-KHz offset step. While in the case of 100-KHz step, the upper limit of the velocity estimation can be extended to be 9000 m/s with a degraded resolution of $1\times10^{-10}$ m/s. Such resolutions can be further enhanced through the use of higher carrier frequency which will be demonstrated in what follows, for a given frequency resolution of the ESA.

We recall that one significant contribution of the proposed approach lies in its frequency-independent operation. Therefore, the proposed estimation approach applied to the DFS estimation for radars and RF communications operating in different frequency bands including for instance from L- to W-bands [5]. Experiments for DFS estimation at other frequency bands will be performed to provide more details. As the frequency of the transmitted signal is specified as 10 GHz, the DFSs are measured by offset steps of 1, 10, and 100 KHz, respectively. The measured DFSs and the related estimation errors are presented in Fig. 7. For another demonstration at 30 GHz, the obtained results are shown in Fig. 8. From the results of the two demonstrations at 10 and 30 GHz, it is clear that estimation errors lower than $\pm5\times10^{-10}$, $\pm5\times10^{-10}$, and $\pm5\times10^{-9}$ Hz are obtained for the three offset steps or the three measurement ranges. Excellent agreements among the results shown in Figs. 5a, 6, 7, and 8, confirm that the implementation of the proposed approach is independent of the frequency or the carrier frequency of the transmitted signals. Therefore, the proposed approach is suitable for DFS estimation in numerous frequency bands (e.g., 10, 15, and 30 GHz in our experiments).

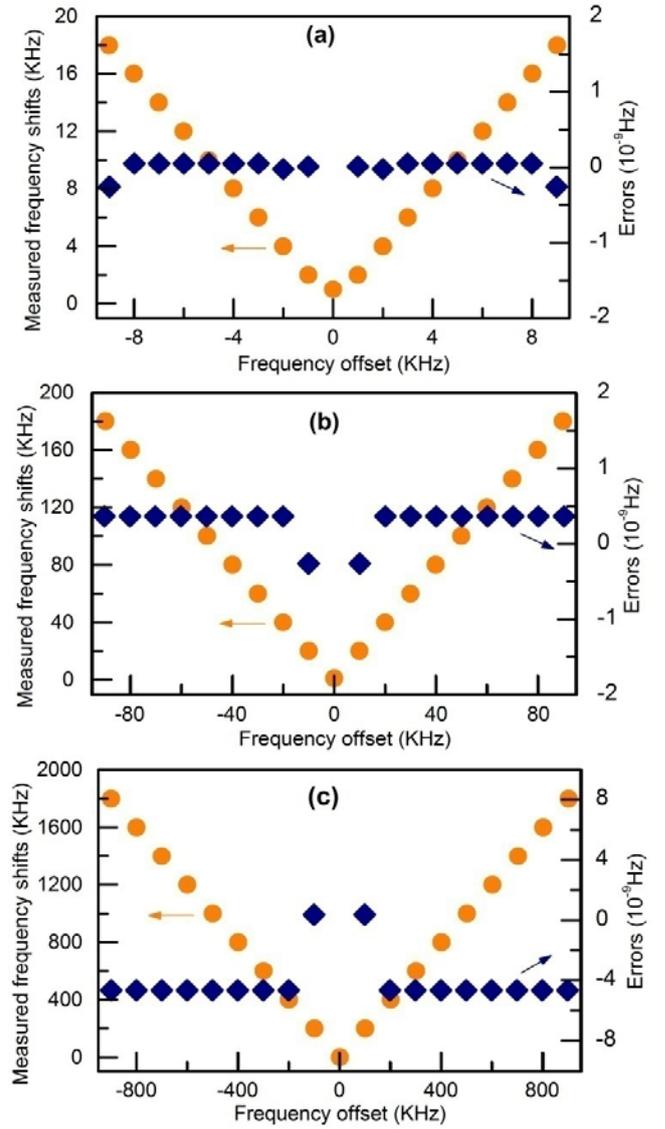

Fig. 7 Measured Doppler frequency shifts and estimation errors for the 10-GHz transmitted signal being tuned at offset steps of (a) 1 KHz, (b) 10 KHz, and (c) 100 KHz.

Here, it should be highlighted that a better resolution or a better capability of detecting low-speed moving object can be realized, as if high-frequency microwave/millimeter-wave signals are used. For instance, the velocity estimation resolution is improved from $1\times10^{-11}$ to $5\times10^{-12}$ m/s, when the carrier frequency is increased from 15 to 30 GHz. In fact, its frequency coverage can be flexibly expanded up to 100 GHz and beyond.

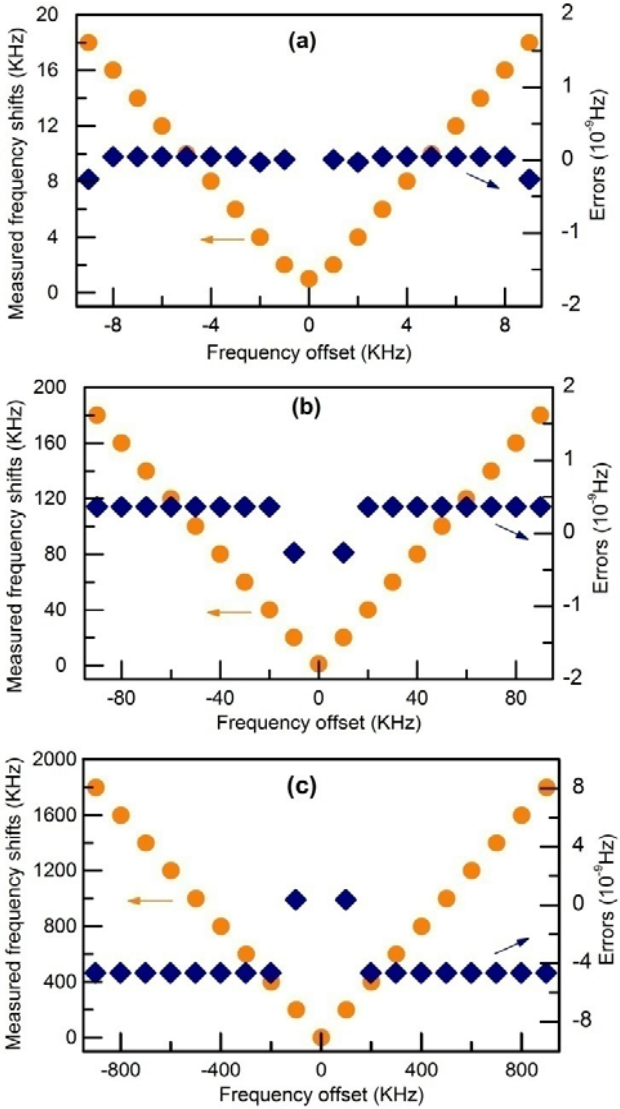

Fig. 8 Measured Doppler frequency shifts and estimation errors for the 30-GHz transmitted signal being tuned at offset steps of (a) 1 KHz, (b) 10 KHz, and (c) 100 KHz.

Some additional insights into the estimation errors obtained at 10, 15, and 30 GHz reveal that the values and the trends are exactly the same for an identical offset step, e.g., 1, 10, or 100 KHz. Demonstrations to clarify this point are listed as follows. For a given step, the DFSs are estimated with the same measurement range, the same RBW, and the same VBW. Meanwhile, the signal to noise ratio (SNR) in the experiments is large enough such that the noise exerts litter influence on the measured electrical spectra which are directly related with the DFSs being estimated. For example, three electrical spectra measured at 10, 15, and 30 GHz are shown in Fig. 9, when the frequency offset is specified as 1 KHz for all the three cases. The three spectra do exhibit obvious differences, but their peak frequencies are exactly the same. That is why identical errors and trends are observed in Figs. 5, 6, 7, and 8 for an identical offset step.

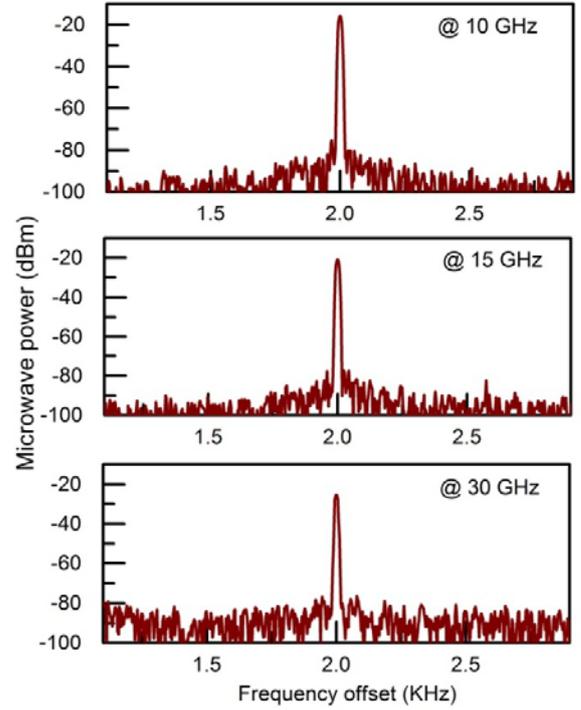

Fig. 9 Measured electrical spectra for the transmitted signal at 10, 15, or 30 GHz as the frequency offset step is specified as 1 KHz.

## 4. Extended discussions

Next, discussions on the advantages and the weakness of the proposed approach are listed in this section. Firstly, the proposed approach is capable of providing a huge wide-open frequency range for DFS estimation, since it is totally independent of the frequency of the transmitted microwave/millimeter-wave signals. This advantage makes it versatile for DFS estimation in a variety of operating frequency bands, by using a single measurement setup. Secondly, the resolutions to DFS estimation and to radial velocity detection have been enhanced by a factor of 2, as a doubled frequency spacing in the optical domain is induced by the DFS to be discriminated. This might be greatly beneficial for the detection of low-speed moving objects and the velocity discrimination requiring finer resolution. In addition, the implementation in the optical domain is completely immune to electromagnetic interference, due to the intrinsic features of photonics.

Furthermore, besides the CW microwave/millimeter-wave signals, the proposed photonic approach also applies to the case of pulsed signals or pulsed Doppler radar. Assume that the pulse repetition frequency is much larger or less than the DFS in specific applications. According to the setup illustrated in Fig. 1, the CW carrier of a pulsed signal can be extracted or provided by the transmitted port to implement the CS-DSB modulation inside EOM-I. While in EOM-II, the pulsed echo signal is applied and a frequency doubled electronic components can be generated to indicate the DFS. Actually, analysis and measurement of pulsed microwave signals have been already realized by employing photonics technology, such as other experiments for instantaneous frequency measurement [33].

On the other hand, the proposed approach is unable to differentiate a negative DFS from a positive one, or a negative radial velocity from a positive one, recently. Both a negative and a positive frequency offsets are converted into a doubled absolute frequency spacing, without any phase information to indicate the direction. As shown in Figs. 5, 7, and 8, only absolute values associated with the DFSs are presented. Thus, to eliminate the direction ambiguity, additional operations or setups should be required, such as the I-Q mixing setup [3]. Regardless of the direction of Doppler effect, the photonic approach provides an effective, powerful solution to DFS estimation and radial velocity measurement, with wide-open frequency range and fine estimation resolution.

## 5. Conclusions

A photonic approach characterized by finer resolution improved by a factor of 2, wide-open frequency range, and immunity to electro-magnetic interference, was proposed and demonstrated for DFS estimation. The DFS estimation was experimentally performed for transmitted microwave/ millimeter-wave signals at 10, 15, and 30 GHz. Within the measurement range from -90 to 90 KHz, the derived DFSs were in good agreement with the theoretical values, with estimation errors lower than $\pm 5 \times 10^{-10}$ Hz. Accordingly, for radial velocity measurement, these results corresponded to a range from 0 to 900 m/s and a resolution of $1 \times 10^{-11}$ m/s at 15-GHz frequency band. This resolution was then enhanced to $5 \times 10^{-12}$ m/s inside the range from 0 to 450 m/s at 30-GHz frequency band. These desirable performances are greatly beneficial for wide-open, high-resolution DFS estimation in radar and microwave/millimeter-wave systems.

## Appendix

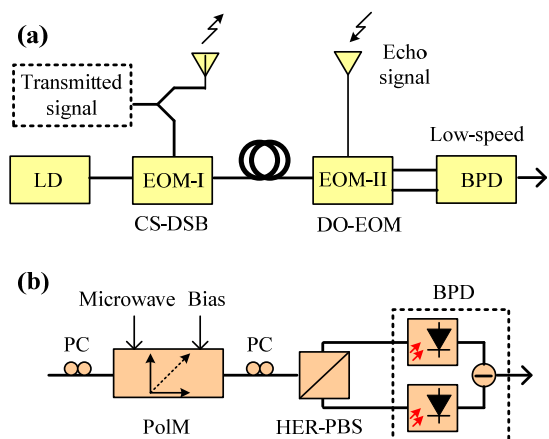

Fig. 10 (a) Schematic diagram of the modified version of the proposed approach and (b) and one example of its implementaiton. (LD, laser diode; EOM, electro-optic modulator; CS-DSB carrier-suppressed double sideband modulation; DO-EOM, dual-output EOM; BPD, balanced photodetector; PC, polarizaiton controller; PolM, polarizaiton modulator; HER-PBS: high-extinction-ratio polarization beam splitter)

An extended version of the proposed approach based on the combination of a dual-output EOM (DO-EOM) and a balanced PD (BPD) is proposed, as shown in Fig. 10. Due to the introduction of balanced detection in the optical domain, the immunity to noise near DC and the sensitivity can be improved. To implement the extended approach, a DO-EOM or an equivalent DO-EOM (e.g., the cascade of a PolM biased at the quadrature point and an HER-PBS in Fig 10b) is needed.